\documentclass[aps,showpacs,showkeys,amsmath,amssymb,floatfix]{revtex4}
\usepackage{graphicx,epsf,subfigure}               % Include figure files
\usepackage{dcolumn}                % Align table columns on decimal point
\usepackage{bm}                     % bold math
\newcommand{\D}{\mathrm{d}}
\begin{document}

\title{Tilt-over mode in a precessing triaxial ellipsoid}

\author{D. C\'ebron}
\email{cebron@irphe.univ-mrs.fr}
\author{M. Le Bars}
\author{ P. Meunier}

\affiliation{Institut de Recherche sur les Ph\'enom\`enes Hors Equilibre, UMR 6594, CNRS et Aix-Marseille Universit\'es, \\
49 rue F. Joliot-Curie, BP146, 13384 Marseille C\'edex 13, France.}

\date{\today}% It is always \today, today,
                 % �but any date may be explicitly specified

\begin{abstract}

The tilt-over mode in a precessing triaxial ellipsoid is studied
theoretically and numerically. Inviscid and viscous analytical
models previously developed for the spheroidal geometry by
Poincar\'e [Bull. Astr. 27, 321 (1910)] and Busse [J. Fluid Mech., 33, 739 (1968)] are extended to this more complex
geometry, which corresponds to a tidally deformed spinning
astrophysical body. As confirmed by three-dimensional numerical
simulations, the proposed analytical model provides an accurate
description of the stationary flow in an arbitrary triaxial
ellipsoid, until the appearance at more vigorous forcing of time
dependent flows driven by tidal and/or precessional instabilities.

\end{abstract}



\maketitle

\section{Introduction} \label{intro}

The flow of a rotating viscous incompressible homogeneous fluid in a
precessing container has been studied for over one century because
of its multiple applications, such as the motions in planetary
liquid cores and the generation of planetary magnetic fields
\cite[e.g.][]{Malkus93,Tilgner_2005}. In the spheroidal geometry,
the early work of \cite{Poincare} demonstrated that the flow of an
inviscid fluid has a uniform vorticity. Later, viscous effects have
been taken into account as a correction to the inviscid modes, in
considering carefully the critical regions of the Ekman layer
\cite[][]{Stewartson_1963,Busse,Greenspan}. Indeed, the Poincar\'e
solution is modified by the apparition of boundary layers, and some
strong internal shear layers are also created in the bulk of the
flow, which do not disappear in the limit of vanishing viscosity
\cite[][]{Busse}. Besides, at high enough precession rates, these
shear layers may become unstable
\cite[][]{malkus68,Vanyo_1995,Noir_2003}, and in a second
transition, the entire flow becomes turbulent: this is the
precession instability. These experimental works have been completed
by numerical studies in cylindrical, spherical and spheroidal
geometries
\cite[][]{Hollerbach_1995,Quartapelle_1995,Tilgner_1999a,Tilgner_2001,Noir_2001},
in particular for studying kinematic dynamo models in a spheroidal
galaxy \cite[][]{Walker_1994}  or for geophysical applications
\cite[][]{Tilgner_1999b,Lorenzani_2001,Lorenzani_2003,Schmitt_2004,Wu_2009}.

\begin{figure}
  \begin{center}
    \epsfysize=7.0cm
    \leavevmode
    \epsfbox{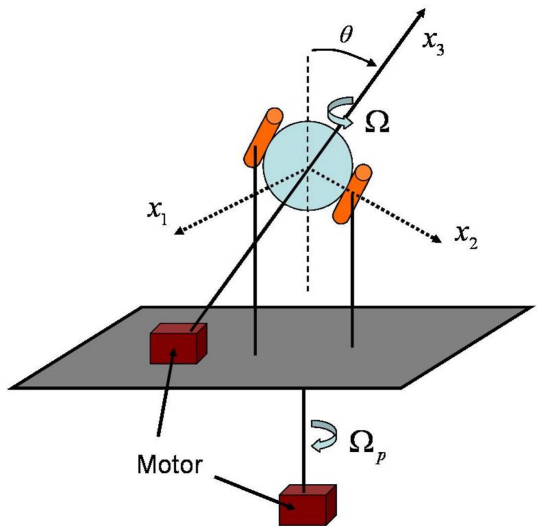}
    \caption{\it{Sketch of the problem under consideration. A
hollow solid but deformable spheroid (i.e. 'the mantle') is filled
with liquid and set in rotation at a constant angular velocity
$\Omega$ versus its axis $(O{x_3})$. The spheroid axis is tilted at
the precession angle $\theta$ and fixed on a rotating table, which
rotates at the precession rate $\Omega_p$. Two fixed rollers aligned
with $(O{x_3})$ then transform the spheroid into a triaxial
ellipsoid by compression along the axis $(O{x_2})$, perpendicular to
the rotation axis. }}
    \label{cebronfig2}
  \end{center}
\end{figure}

However, in natural systems, both the rotation and the gravitational
tides deform the celestial body into a triaxial ellipsoid, where the
so-called elliptical (or tidal) instability may take place (see
\cite[][]{KerswellMalkus,Lacaze_2004,LeBars09,Cebron_2010,Kerswell_2002}
for details on this instability and its geo- and astro-physical
applications). The elliptical instability takes place in any
rotating fluid whose streamlines are elliptically deformed (see e.g.
\cite{Waleffe_1990} or \cite{LeBars07}). It comes from a parametric
resonance of two inertial waves of the rotating fluid with the tidal
(or elliptical) deformation of azimutal wavenumber $m=2$
\cite[][]{Kerswell_2002}. Similarly, it has been suggested that the
precession instability comes from the parametric resonance of two
inertial waves with the forcing related to the precession of
azimutal wavenumber $m=1$, which comes from the deviations of the
laminar tilt-over base flow from a pure solid body rotation
\cite[][]{Kerswell_1993,Kerswell_1995,Lagrange2008}. But it has also
been suggested that the precession instability is related to a shear
instability of the zonal flows that appear in a precessing container
\cite[e.g.][]{Lorenzani_2003}. Clearly, the precise origin of the
precession instability is still under debate, and is beyond the
point of the present work. But since tides and precession are
simultaneously present in natural systems, it seems necessary to
study their reciprocal influence, in presence or not of
instabilities. The full problem is rather complex and involves three
different rotating frames: the precessing frame, with a period $T_p
\approx 26 000$ years for the Earth, the frame of the tidal bulge,
with a period around $T_d \approx 27$ days for the Earth, and the
container or 'mantle' frame, with a period $T_s \approx 23.93$ hours for the
Earth. As a first step towards the full study of the interaction
between the elliptical instability and the precession, we consider
the particular case where the triaxial ellipsoid is fixed in the
precessing frame ($T_d=T_p$), which allows the theoretical approach
to be analytically tractable. We show in figure \ref{cebronfig2} a
sketch of this configuration.

The paper is organized as follow. In section \ref{triaxial},
Poincar\'e and Busse analytical models are extended to precessing
triaxial ellipsoids. Then in section \ref{validation}, our analysis
is validated by comparison with a numerical simulation.

\section{Analytical solution of the flow in a precessing triaxial ellipsoid}\label{triaxial}

In this section, we consider firstly the case of an inviscid fluid
and extend the Poincar\'e model to a precessing triaxial ellipsoid,
which allows us to obtain explicit analytical solutions. We then
tackle the viscous case in extending the Busse model, following the
method of \cite{Noir_2003}. As sketched in figure \ref{cebronfig2},
we consider the rotating flow inside a precessing triaxial
ellipsoidal container of principal axes $(a_1,a_2,a_3)$. We define
$(Ox_1,Ox_2,Ox_3)$ in the frame of the tidal bulge which is also the
precessing frame, such that $Ox_i$ is along the principal axis $a_i$
of the ellipsoidal container and $(O{x_3})$ is the mantle rotation
axis. We note $\Omega$ the imposed mantle angular velocity and use
$\Omega^{-1}$ as a timescale. We also introduce the mean equatorial
radius $R_{eq}=(a_1+a_2)/2$, which is used as a lengthscale.
Consequently, the problem is fully described by six dimensionless
numbers: the eccentricity $\epsilon=(a_1^2-a_2^2)/(a_1^2+a_2^2)$,
the aspect ratio $a_3/a_1$, the Ekman number $E=\nu / \Omega
R_{eq}^2$, where $\nu$ is the kinematic viscosity of the fluid, and
the three components of the dimensionless precession vector
$\boldsymbol{\Omega_p}$ in the inertial frame of reference, i.e. the
angle $\theta$ between $\boldsymbol{\Omega_p}$ and
$\boldsymbol{e_{x_3}}$, the angle $\theta_2$ between
$\boldsymbol{\Omega_p}$ and $\boldsymbol{e_{x_1}}$ and the angular
precession rate $\Omega_p$, which is positive for prograde
precession and negative for retrograde precession.

\subsection{Inviscid Poincar\'e tilt-over mode for a triaxial ellipsoid}\label{inviscid}

We look for a base flow solution of the Euler equations, i.e.
\begin{eqnarray}\label{euler}
\frac{\partial \mathbf{u}}{\partial t}+ \mathbf{u} \cdot
\mathbf{\nabla} \mathbf{u} = -\mathbf{\nabla} p - 2
\boldsymbol{\Omega_p} \times \boldsymbol{u},
\end{eqnarray}
\begin{eqnarray}
\mathbf{\nabla}  \cdot \mathbf{u} =0,
\end{eqnarray}
inside the triaxial ellipsoid in the precessing frame. In this
frame, using the ellipsoid equation
\begin{eqnarray}
\frac{x_1^2}{a_1^2}+\frac{x_2^2}{a_2^2}+\frac{x_3^2}{a_3^2}=1,
\end{eqnarray}
we transform the ellipsoidal geometry into a sphere with the
transformation $(x'_k)=\left( \frac{x_k}{a_k} \right)_{k \in
(1,2,3)}$ and we write the velocity field in this sphere as
\begin{eqnarray}
(U'_k)=\left( \frac{U_k}{a_k}\right)_{k \in (1,2,3)}.
\label{transform}
\end{eqnarray}

As suggested in \cite{Poincare}, we focus on the so-called 'simple
motions' such that the velocity $\mathbf{U}(U_1,U_2,U_3)$ is
described as linear combinations of the coordinates. This hypothesis
leads to a solid body rotation in the sphere, i.e.
$\mathbf{U'}=\boldsymbol{\omega'} \times \mathbf{r'}$, where
$\boldsymbol{\omega'}(\omega'_1,\omega'_2,\omega'_3)$ depends on
time a priori. Equation \ref{transform} then gives the following
velocity field in the ellipsoid:
\begin{eqnarray}
\mathbf{U}= \left( U_k \right)_{k \in (1,2,3)} = \left( \frac{a_{k}}{a_{k-1}}\ \omega'_{k+1} x_{k-1} -
\frac{a_k}{a_{k+1}}\ \omega'_{k-1} x_{k+1} \right)_{k \in (1,2,3)}
\label{vitesse},
\end{eqnarray}
where permutations ${k \in (1,2,3)}$ are used.

Now, we have to find the 'simple motions' solution of the Euler
equations for the rotational part
of the flow, taking into account the no-penetration %stress
 boundary conditions for its
irrotational part, which leads to the so-called Poincar\'e flow. We
follow the method of \cite{Noir_2000} rather than the lagrangian
method of Poincar\'e which is more laborious. We deduce the rotation
rate vector $\boldsymbol{\omega}$ from the velocity field
(\ref{vitesse}):
\begin{eqnarray}
\boldsymbol{\omega}=\frac{1}{2}\ \mathbf{\nabla} \times
\boldsymbol{U}=
   \left (
       \omega_k
   \right )_{k \in (1,2,3)}=\frac{1}{2}
   \left (
       \left ( \frac{a_{k-1}}{a_{k+1}} + \frac{a_{k+1}}{a_{k-1}} \right )\ \omega'_k
   \right )_{k \in (1,2,3)}.
\end{eqnarray}
Note that $\boldsymbol{\omega'}$ is independent of the space
coordinates so that $\boldsymbol{\omega}$ is uniform. Taking the
rotational of (\ref{euler}), the inviscid equation for
$\boldsymbol{\omega}$ gives the following three scalar equations:
\begin{eqnarray}
\frac{\D \omega_k}{\D t}+ ( \alpha_{k+1,k} -
\alpha_{k-1,k})\ \omega_{k-1}\ \omega_{k+1} = \alpha_{k-1,k}\
\Omega_{p,k-1}\ \omega_{k+1}-\alpha_{k+1,k}\ \Omega_{p,k+1}\
\omega_{k-1} \label{eq:system}
\end{eqnarray}
for permutations  $k \in (1,2,3)$, with the coefficients
$\alpha_{i,j}= \frac{2}{\eta_{ij}+2} = 2\
\frac{\eta_{ji}+1}{\eta_{ji}+2}$ and the different ellipticities
$\eta_{ij}=\frac{a_i^2-a_j^2}{a_j^2}$ of the container. The
equations for a spheroidal geometry are recovered with $a_1=a_2$.

We focus on the stationary solutions of the problem. In this
particular case, we solve the system (\ref{eq:system}) analytically,
which gives:
\begin{eqnarray}
\omega_1=  \frac{\omega_3}{\beta_{12}}
\frac{a_3^2+a_2^2}{\gamma_{1}}\ \Omega_{p,1} \label{eq:om1},
\end{eqnarray}
\begin{eqnarray}
\omega_2= \frac{\omega_3}{\beta_{12}} \frac{a_3^2+a_1^2}{\gamma_{2}}\
\Omega_{p,2}, \label{eq:om2}
\end{eqnarray}
where $\beta_{ij}=\frac{a_i^2+a_j^2}{2a_i a_j}=\beta_{ji}$ and
$\gamma_i=\frac{\omega_3}{\beta_{12}} (a_i^2-a_3^2)+2\ \Omega_{p,3}\
a_1\ a_2$. The corresponding velocity field writes:
\begin{eqnarray}
U_1= a_1\ \frac{\omega_3}{\beta_{12}} \left( -\frac{x_2}{a_2}+\frac{2\
a_1\ \Omega_{p,2} x_3}{\gamma_2} \right), \label{U1}
\end{eqnarray}
\begin{eqnarray}
U_2= a_2\ \frac{\omega_3}{\beta_{12}} \left( \frac{x_1}{a_1}-\frac{2\
a_2\ \Omega_{p,1} x_3}{\gamma_1} \right),\label{U2}
\end{eqnarray}
\begin{eqnarray}
U_3= a_3\ \frac{\omega_3}{\beta_{12}} \left( \frac{2\ a_3\ \Omega_{p,1}
x_2}{\gamma_1}-\frac{2\ a_3\ \Omega_{p,2} x_1}{\gamma_2} \right). \label{U3}
\end{eqnarray}
Note that the choice of $\omega_3$ is arbitrary here. Actually,
$\omega_3$ is determined by the boundary (Ekman) layer and thus has
to be determined by a viscous study, following for instance the
method of \cite{Busse}, as described in the next section. Note also
that this velocity field is divergent for $\gamma_1=0$ or
$\gamma_2=0$: the inviscid study gives two resonances, which correspond to resonance between the frequencies of respectively the precessional forcing and the tilt-over (see \cite{Noir_2003} for details). These linear
resonances are reached for two specific precession rates (depending
on the aspect ratio) which are
\begin{eqnarray}
\Omega_{p,3}= \frac{a_3^2-a_i^2}{a_1^2+a_2^2}\ \omega_3 \label{eq:resonance}
\end{eqnarray}
for $i \in (1,2)$. It is clear that oblate ellipsoids
($a_1,a_2>a_3$) have their resonance in the retrograde regime (i.e.
in the range $\Omega_{p,3}<0$) whereas prolate ellipsoids
($a_1,a_2>a_3$) have their resonance in the prograde regime.
Finally, compared to the spheroidal case, an important result here
is the apparition of the second resonance created by the equatorial
ellipticity of the container.

\subsection{Viscous study of the tilt-over mode in a triaxial ellipsoid}\label{viscous}

Following \cite{Busse}, it is possible to take into account the
viscosity in the study of the flow in a precessing triaxial
ellipsoid. Here, we focus on the equivalent method of
\cite{Noir_2003}, based on the equilibrium between the inertial
torque $\boldsymbol{\Gamma_i}$, the pressure torque
$\boldsymbol{\Gamma_p}$, and the viscous torque
$\boldsymbol{\Gamma_v}$. Keeping the leading terms for these three
torques, the general torque balance given in \cite{Busse} for a
steady rotating flow $\boldsymbol{q}=\boldsymbol{\omega}\times
\boldsymbol{r}$ in the precessing frame within a volume $V$ with a
surface $\Sigma$ writes:
\begin{eqnarray}
\overbrace{2 \int_V \boldsymbol{r} \times ( \boldsymbol{\Omega_p}
\times \boldsymbol{q}) dV}^{\boldsymbol{\Gamma_i}} = \overbrace{-
\oint_\Sigma p \boldsymbol{r} \times \boldsymbol{n}
d\Sigma}^{\boldsymbol{\Gamma_p}} + \overbrace{E \int_V
\boldsymbol{r} \times \nabla^2 \boldsymbol{q}
dV}^{\boldsymbol{\Gamma_v}} \label{torque},
\end{eqnarray}
where $\boldsymbol{r}$ is the position vector and $\boldsymbol{n} $
is the unit vector normal to $\Sigma$ pointing outward
\cite[see][]{Noir_2003}. Now, we have to calculate these terms for a
triaxial ellipsoid.

At first order, the pressure gradient equilibrates the centrifugal
force, which gives:
\begin{eqnarray}
p=\frac{1}{2} \sum\limits_{\substack{i=1
}}^{3}{\left[(\omega_{i+1}^2+\omega_{i-1}^2)\
x_i^2-\sum\limits_{\substack{j=1 \\ j \neq i}}^{3}{\omega_i\ \omega_j\
x_i\ x_j}\right]}.
\end{eqnarray}
Hence in the limit of small ellipticities, i.e. at first order in
$\eta=1-a_3/a_1$ and $\eta_2=1-a_2/a_1$, the pressure torque writes:
\begin{eqnarray}
\boldsymbol{\Gamma_p}= - \oint p\ \boldsymbol{r} \times \boldsymbol{n}\ \D \Sigma = \left( \Gamma_{p,k} \right)_{k \in (1,2,3)} = I
   \left (
   \begin{array}{ccc}
       (\eta - \eta_2)\ \omega_2\ \omega_3 \\
       - \eta\ \omega_1\ \omega_3 \\
       \eta_2\ \omega_1\ \omega_2 \\
   \end{array}
   \right ), \label{torque_p}
\end{eqnarray}
where I is the moment of inertia in the spherical approximation.
Note that the expression of \cite{Noir_2003} is recovered for the
spheroid (i.e. $a_1=a_2$ hence $\eta_{2}=0$).

After little algebra, the precessional torque simply writes:
\begin{eqnarray}
\boldsymbol{\Gamma_i}= I \boldsymbol{\Omega_p} \times
\boldsymbol{\omega} \label{torque_i}.
\end{eqnarray}

Finally, the equations (\ref{torque_p}) and (\ref{torque_i}) give
$\boldsymbol{\Gamma_p} \cdot \boldsymbol{\omega}=0 $ and
$\boldsymbol{\Gamma_i} \cdot \boldsymbol{\omega}=0 $, and thus with
the equation (\ref{torque}) $\boldsymbol{\Gamma_v} \cdot \boldsymbol{\omega}=0$.
Consequently, the fluid being in a stationary state, there is no
differential rotation along $\boldsymbol{\omega}$:
\begin{eqnarray}
\boldsymbol{\omega} \cdot \boldsymbol{e_3}=\omega^2.
\label{eq:nospinup}
\end{eqnarray}
Indeed, in the rotating frame of the fluid, the angular rate of the
container along $\boldsymbol{\omega}$ is null: the only differential
rotation between the fluid and the container is in the equatorial
plane such that no spin-up process occurs. Note that this can also
be recovered with a boundary layer analysis, in the same way as
\cite{Busse}, which shows that the volumic Ekman pumping is solution
of the inviscid bulk equations provided that the so-called
solvability condition (\ref{eq:nospinup}) is verified. According to
this equation, also named the no spin-up condition in
\cite{Noir_2003}, the only relative motion between the interior and
the boundary is a rotation given by
$\boldsymbol{\omega_{eq}}=\boldsymbol{\omega}-\boldsymbol{\Omega}$.
Then, we need to calculate the viscous torque due this equatorial
differential rotation. This calculation relies on the fact that in
order to maintain the basic stationary state, the torque supplied to
the fluid to counter balance the viscous friction is given by the
decay rate that would occur if at a given time the precession is
turned off. This decay rate is simply given by the rate at which
energy is dissipated by the Poincar\'e mode in a free system at
$t=0$. This rate is given by the Greenspan's theory
\cite[][]{Greenspan}, valid in the frame rotating with the fluid.
Thus, this linear solution for the viscous decay of the spin-over
mode in a rotating fluid leads to introduce a new Ekman number
$E_f=E/ \omega$ and a new unit of time $\tilde{t}=t \omega$, scaled
with the fluid rotation rate $\omega$. According to
\cite{Greenspan}, the time evolution of $\boldsymbol{\omega_{eq}}$
in the non-rotating frame is:
\begin{eqnarray}
\boldsymbol{\omega_{eq}}(\tilde{t})= e^{\lambda_r\ \tilde{t}\  \sqrt{E_f}
} \left[ \cos \left(\lambda_i\
\tilde{t}\  \sqrt{E_f} \right)\boldsymbol{\omega_{eq}}(0)-\sin \left(\lambda_i\ \tilde{t}\ \sqrt{E_f}
 \right) \frac{\boldsymbol{\omega}\times
\boldsymbol{\omega_{eq}}(0)}{\omega} \right]
\end{eqnarray}
with $\lambda_r=-2.62$ and $\lambda_i=0.259$. Note that strictly
speaking in triaxial ellipsoids, the ellipticity modify the growth
rate and eigenfrequency of inertial modes. However, in the limit of
small ellipticities we consider here, this modification can be
neglected.

Using this equation, reintroducing the variables $E$ and $t$, the
equatorial viscous torque is:
\begin{eqnarray}
\boldsymbol{\Gamma_v}= I \left( \frac{\D \boldsymbol{\omega_{eq}}}
{\D t} \right)_{t=0}=I \sqrt{\omega E}
   \left (
   \begin{array}{ccc}
       \lambda_r\ \omega_1 + \lambda_i\ \omega_2 / \omega \\
       \lambda_r\ \omega_2 - \lambda_i\ \omega_1 / \omega \\
       \lambda_r\ (\omega_3-1) \\
   \end{array}
   \right )
\end{eqnarray}

Then, the torque balance given by (\ref{torque}) projected onto the
rotation axis of the fluid $\boldsymbol{\omega}$ (the no spin-up
condition (\ref{eq:nospinup})), as well as onto the principal axes
$\boldsymbol{e_{x_1}}$ and $\boldsymbol{e_{x_3}}$ yields the
following system of equations:
\begin{eqnarray}
   \omega_1^2+\omega_2^2=\omega_3\ (1-\omega_3)
\end{eqnarray}
\begin{eqnarray}
   \Omega_{p,2}\ \omega_3 - \Omega_{p,3}\ \omega_2=  (\eta - \eta_2)\ \omega_2\ \omega_3 + \left( \lambda_r\ \omega_1\ \omega_3^{1/4}+\lambda_i\ \frac{\omega_2}{\omega_3^{1/4}}  \right)\sqrt{E}
\end{eqnarray}
\begin{eqnarray}
   \Omega_{p,1}\ \omega_2 - \Omega_{p,2}\ \omega_1=  \eta_2\ \omega_1\ \omega_2 -\lambda_r\ \omega_3^{1/4}\
   (1-\omega_3)\ \sqrt{E}.
\end{eqnarray}

\begin{figure}
  \begin{center}
    \epsfysize=7.0cm
    \leavevmode
    \epsfbox{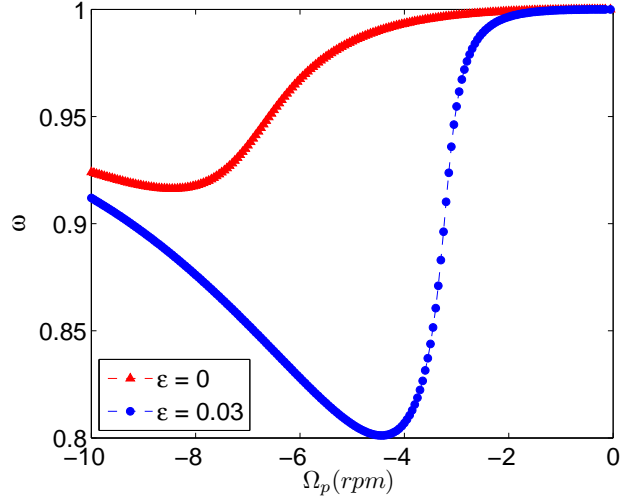}
    \caption{\it{Theoretical amplitude of tilt-over mode angular rotation rate $\omega$ for a spheroid ($\epsilon=0$) and for a slightly deformed triaxial ellipsoid ($\epsilon=0.03$). The other parameters used in this figure are those of \cite{Noir_2003}: the rotation rate of the container $\Omega=207\ rpm$ and $a_1=0.125\ m$ (such that the Ekman number is $E=3 \cdot 10^{-5}$), $\theta=9^{\circ}$, $a_3/a_1=0.96$ and $\theta_2=0^{\circ}$. The spheroidal case corresponds to the plot $\nu=10^{-5}\ m^2/s$ in figure 3 of \cite{Noir_2003}. Note that even a very small tidal deformation significantly changes the obtained solution.}}
    \label{cebronfig2bis}
  \end{center}
\end{figure}

The supplementary terms compared to \cite{Busse} or \cite{Noir_2003}
do not allow to simplify the system of equations into only one, and
the full system has to be solved numerically to obtain the rotation
axis components of the fluid. This non-linear system can be solved
in an efficient way with a continuation method (successive
perturbations on the $a_2$ axis) starting from the Busse's solution
in a spheroid. An example is shown in figure \ref{cebronfig2bis}
where the solution in the spheroidal geometry (the case
$\nu=10^{-5}\ m^2/s$ in figure 3 of \cite{Noir_2003}, which gives a
Ekman number of $E=3 \cdot 10^{-5}$) is compared to a slightly
deformed triaxial ellipsoid ($\epsilon=0.03$) with the same ratio
$a_3/a_1$. It can be noticed that even a very small tidal
deformation $\epsilon$ radically changes the obtained solution.

\section{Numerical and experimental validation}\label{validation}

Our purpose here is to validate and test the range of validity of
our analytical solution by comparison with numerical simulations of
the full non-linear Navier-Stokes equations in a precessing
ellipsoid.

\subsection{Numerical resolution}

We consider the rotating flow inside a triaxial ellipsoidal
container of principal axes $(a_1,a_2,a_3)$, as sketched in figure
\ref{cebronfig2}. We work in the precessing frame of reference.
Starting from rest, a constant tangential velocity
$U\sqrt{1-(x_3/a_3)^2}$ is imposed from time $t=0$ all along the
outer boundary in each plane of coordinate $x_3$ perpendicular to
the rotation axis $(O{x_3})$, where $U$ is the imposed boundary
velocity at the equator. We introduce the timescale $\Omega^{-1}$ by
writing the tangential velocity along the deformed outer boundary at
the equator $U=\Omega R_{eq}$. We then solve the Navier-Stokes
equations with no-slip boundary conditions, taking into account a
Coriolis force associated with the precession
$\boldsymbol{\Omega_p}$, i.e. in the frame $(Ox_1,Ox_2,Ox_3)$ of the
tidal bulge which is also the precessing frame, we solve:
\begin{eqnarray}
\frac{\partial \mathbf{u}}{\partial t}+ \mathbf{u} \cdot
\mathbf{\nabla} \mathbf{u} = -\mathbf{\nabla} p + E
\boldsymbol{\bigtriangleup} \mathbf{u} - 2 \boldsymbol{\Omega_p}
\times \boldsymbol{u},
\end{eqnarray}
\begin{eqnarray}
\mathbf{\nabla}  \cdot \mathbf{u} =0.
\end{eqnarray}

In this work, the range of parameters studied is $E \geq 10^{-3}$
and $\epsilon \leq 0.32$. Once a stationary or periodic state is
reached, we determine the rotation rate in the bulk of the fluid,
i.e. outside the viscous boundary layer. To do so, we introduce an
interior homothetic ellipsoid, in a ratio $\kappa$, and we define
the bulk rotation rate $\boldsymbol{\omega}$ as the mean value of
rotation rate over this homothetic ellipsoid. Following
\cite{Owen_1989}, we consider a dimensionless viscous layer
thickness of $\delta_{\nu} \approx 5 \sqrt{E}$ and thus choose for the homothetic ratio $\kappa \approx 1-\delta_{\nu} \approx 1-5
\sqrt{E}$. Note that the spheroidal case can be efficiently solved
by spectral methods (see e.g. \cite{Schmitt_2004} or
\cite{Wu_2009}). But for the triaxial ellipsoids we are interested
in, there is no simple symmetry. Our computations are thus performed
with a finite element method, which allows us to correctly reproduce
the geometry and to simply impose the boundary conditions. The
solver and the numerical method are described in details in
\cite{Cebron_2010}.

\subsection{Experimental set-up}

The experimental set-up has been described in \cite{Meunier2008,Lagrange2008} for a precessing
cylinder and readers should refer to these papers for more details. For this paper, the experiment
has been slightly modified in order to study the precession of a spheroid. This allows to validate
the flow at small Ekman numbers, which is impossible numerically. Unfortunately, this set-up is
limited to a spheroidal geometry ($a_1=a_2$) and is not able to validate the theory for a
tri-axial ellipsoid. Further modifications to the set-up, such as two rollers compressing the
spheroid, would be needed to make the equatorial plane elliptical.

The home-made spheroid has been obtained by assembling two half-spheroidal cavities drilled in
solid Plexiglas cylinders. The accuracy of the machines ($10$ microns) ensured that the step between
the two parts would be smaller than the Ekman layer (of the order of $300$ microns at $E=10^{-5}$).
This spheroid of equatorial diameter $17$~cm and aspect ratio $a_3/a_1=0.85$ is filled with water
and mounted on a motor which is itself located on a rotating platform. The angular velocities of
the spheroid and the platform are stable within $0.1\%$ and the precessing angle $\theta$ between
the two axes  was varied from $5^{\circ}$ to $15^{\circ}$ with an accuracy of $0.1^{\circ}$.

Pulsed Yag lasers are used to create a luminous sheet perpendicular to the axis of the rotating
platform. A PIV (Particle Image Velocimetry) camera is located on the rotating platform, aligned with the axis of rotation of
the spheroid. This allowed to obtain PIV measurements in a plane almost parallel to the equatorial
plane $(x_1,x_2)$ and located at a distance $x_3=5$ cm above it. Since the Yag lasers are not
located on the rotating platform, the measurement plane is tilted with an angle $\theta$ with
respect to the equatorial plane, which introduces an error of up to 2cm (for $\theta=15^{\circ}$)
in the axial location of the velocity vectors. However, the measured velocity components exactly
correspond to the $x_1,x_2$ components of the velocity because the camera is aligned with the axis
of the spheroid. There is no distortion of the images at the air-Plexiglas interface (because it
is a plane) but there are distortions of the images due to the Plexiglas-water spheroidal
interface. These deformations are small because the refractive index of the Plexiglas and the
water are close. They were calculated analytically and checked experimentally using a grid. They
are located mostly at the boundary of the spheroid and they introduce a maximum error of $15\%$ on
the radial displacement of the particles. This does not bias the measurements because only the
central region of the velocity field was used for the treatment of the data. The PIV images were
rotated numerically in order to remove the background rotation of the flow before being treated by
a home-made cross-correlation PIV algorithm.

In the frame of reference of the spheroid, the 2D velocity field was found to be a nearly uniform
translation flow whose direction and amplitude vary with the precession frequency $\Omega_p$. This
mean flow is linked to the equatorial component of the angular rotation rate $(\omega_1,\omega_2)$
which creates a uniform translation flow $(\alpha_{31}\ x_3\ \omega_2,-\ \alpha_{32}\ x_3\
\omega_1)$ in the plane $x_3=5$ cm. This flow corresponds to the last terms found in equations
(\ref{U1}) and (\ref{U2}) in the case $a_1=a_2$. The first terms of these equations correspond to the
axial angular rotation $(\omega_3)$, which in the frame of reference of the spheroid is equal to
$\omega_3-1$ in dimensionless form and is thus small outside of the resonances. The measurements
of the mean velocity and of the mean vorticity thus give in a simple manner the three components
of the rotation vector $\boldsymbol{\omega}$. Such measurements have been done for the first time
in a precessing spheroid, and will be compared to theoretical and numerical results in the
following.

\subsection{Validation in the spheroidal case}\label{valid}

Before dealing with the problem of precessing triaxial ellipsoids, a
first step of this work has been to simultaneously check the
validity of our numerical tool and the validity of the theoretical
development of \cite{Busse} over an extended range in Ekman number.
To do so, we combine numerical and experimental approaches.

\begin{figure}
  \begin{center}
    \begin{tabular}{ccc}
      \setlength{\epsfysize}{9.0cm}
      \subfigure[]{\epsfbox{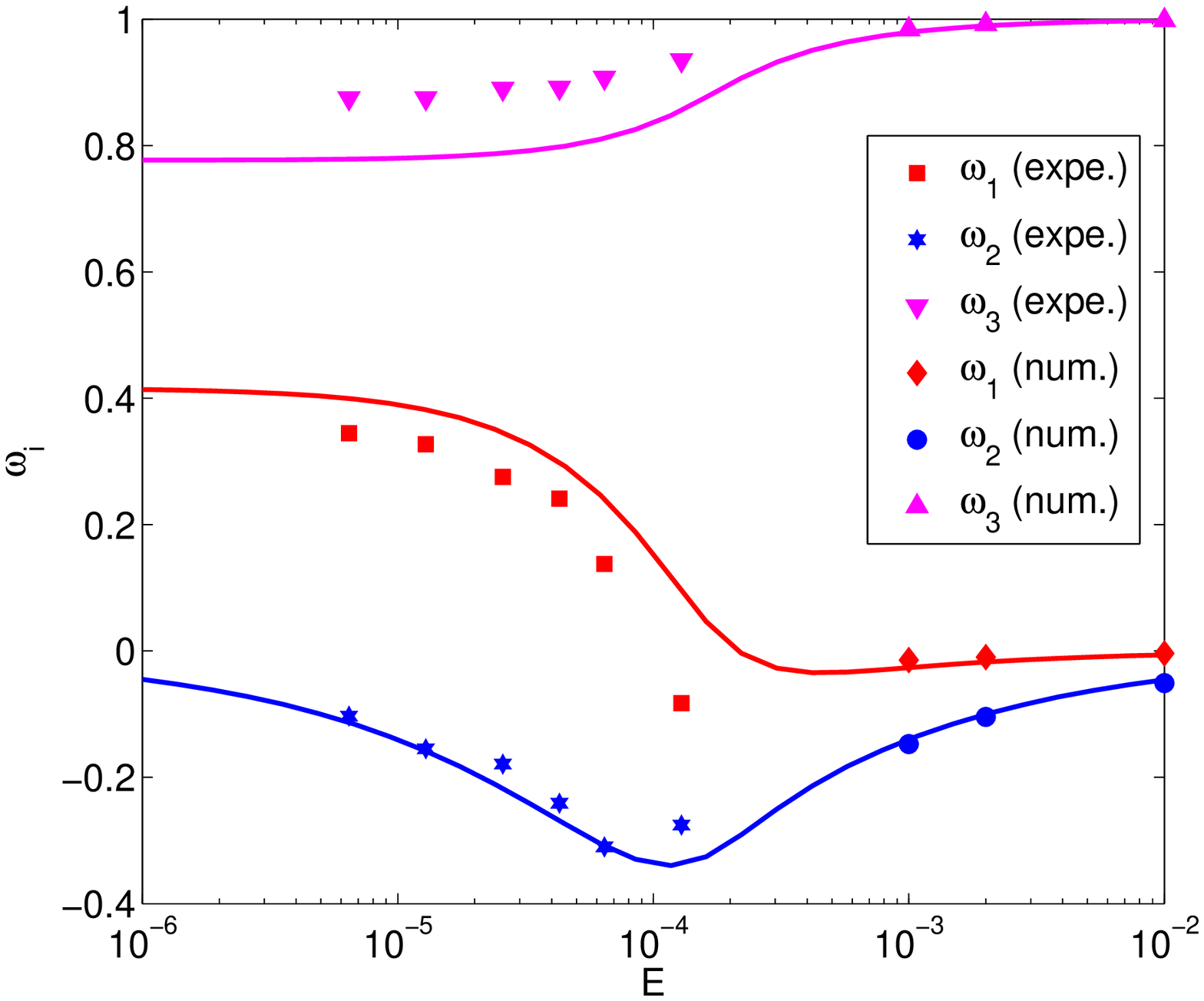}} \\
      \setlength{\epsfysize}{9.0cm}
      \subfigure[]{\epsfbox{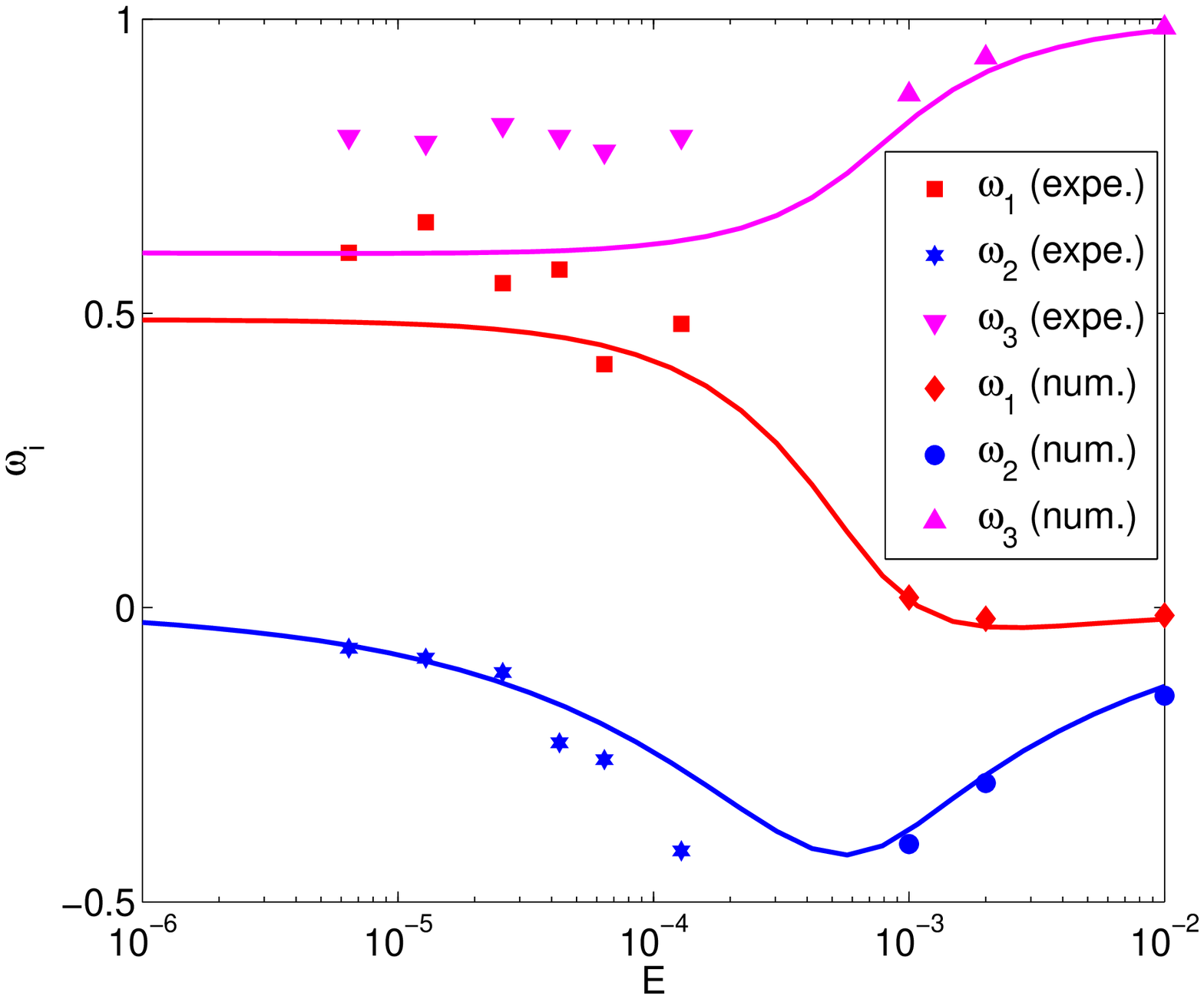}}
     \end{tabular}
\caption{\it{Comparison of the theoretical rotation rate components
of Busse's solution for a spheroid (continuous lines) with
experimental PIV measurements and numerical simulations performed
over a large range of Ekman numbers for $\Omega_p=-0.14$ and
$a_3/a_1=0.85$. (a) For a precession angle of $5^{\circ}$. (b) For a
precession angle of $15^{\circ}$.}}
    \label{cebronfig1}             % Pensez ? mettre le nom du premier auteur ? la place de Nom
  \end{center}
\end{figure}

In figure \ref{cebronfig1}, the theory proposed in \cite{Busse} is
validated over more than three decades of Ekman numbers for two
different angles of precession. The interest of the presented
results is twofold. Firstly, they validate our numerical model,
which will now be used to study triaxial ellipsoids. Secondly, this
completes the previous validations of Busse's theory \cite[e.g.
][]{Tilgner_2001,Noir_2003}, in particular for rather large Ekman
numbers. Note that, as already noted in \cite{Tilgner_2001}, the
theoretical analysis requires that the angle between the rotation
vector of the container and the rotation vector of the fluid is
small since otherwise the Ekman boundary layer analysis is no longer
valid. This could explain the differences between the experimental
determinations and the theoretical predictions at the highest
precessional forcing angle ($\theta=15^{\circ}$). A supplementary
confirmation of both the numerical model and Busse's theory at such
Ekman numbers is given in figure \ref{cebronfig12} (a) where the
components of the tilt-over rotation axis are in an excellent
agreement for a large range of precession rate. At large precession
rates, the inviscid results of \cite{Poincare} are recovered.

\subsection{Numerical simulations of precessing triaxial ellipsoids}

\begin{figure}                  % Chaque figure doit avoir pour nom nomfig1.eps,
  \begin{center}
    \begin{tabular}{ccc}
      \setlength{\epsfysize}{7.0cm}
      \subfigure[]{\epsfbox{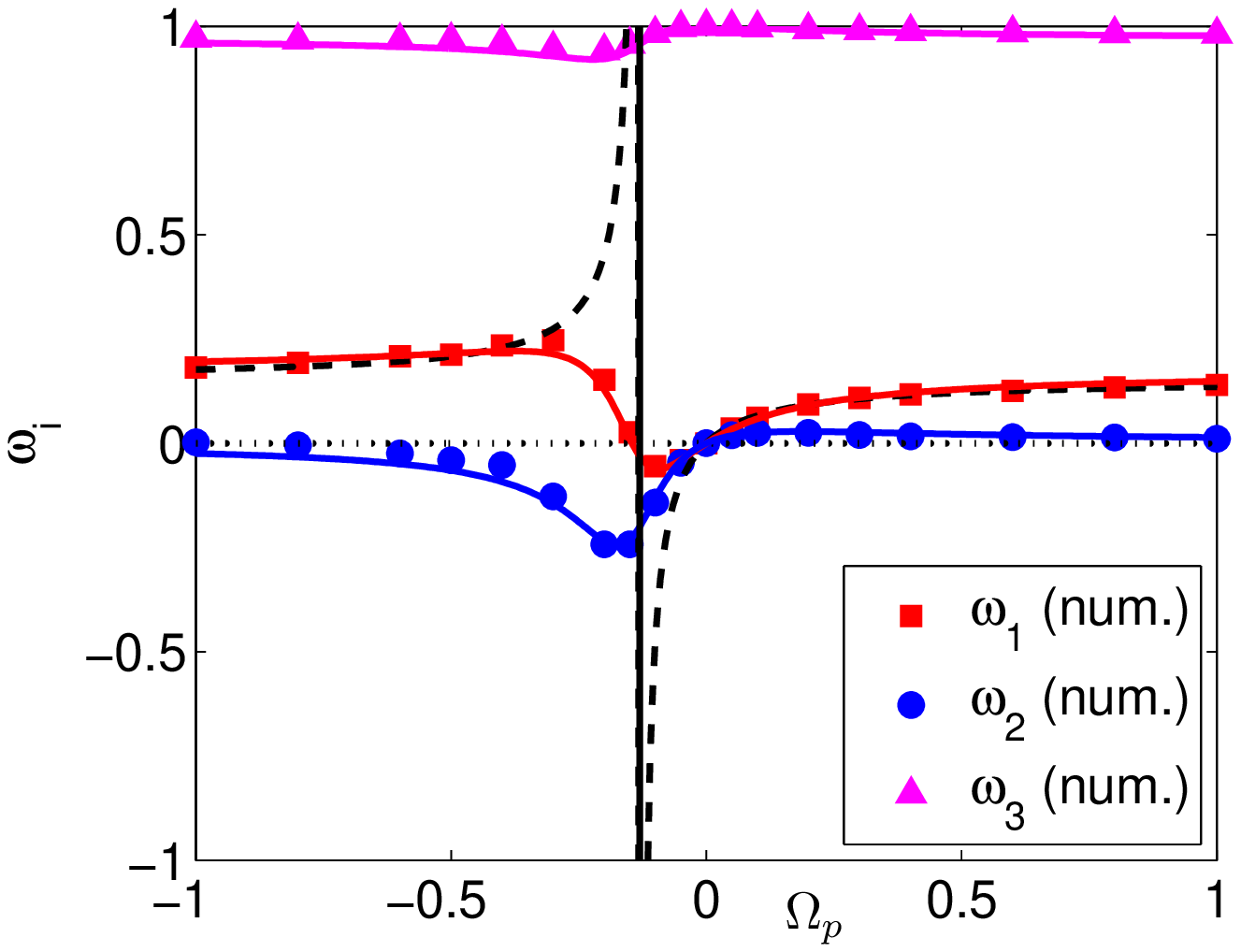}} \\
      \setlength{\epsfysize}{7.0cm}
      \subfigure[]{\epsfbox{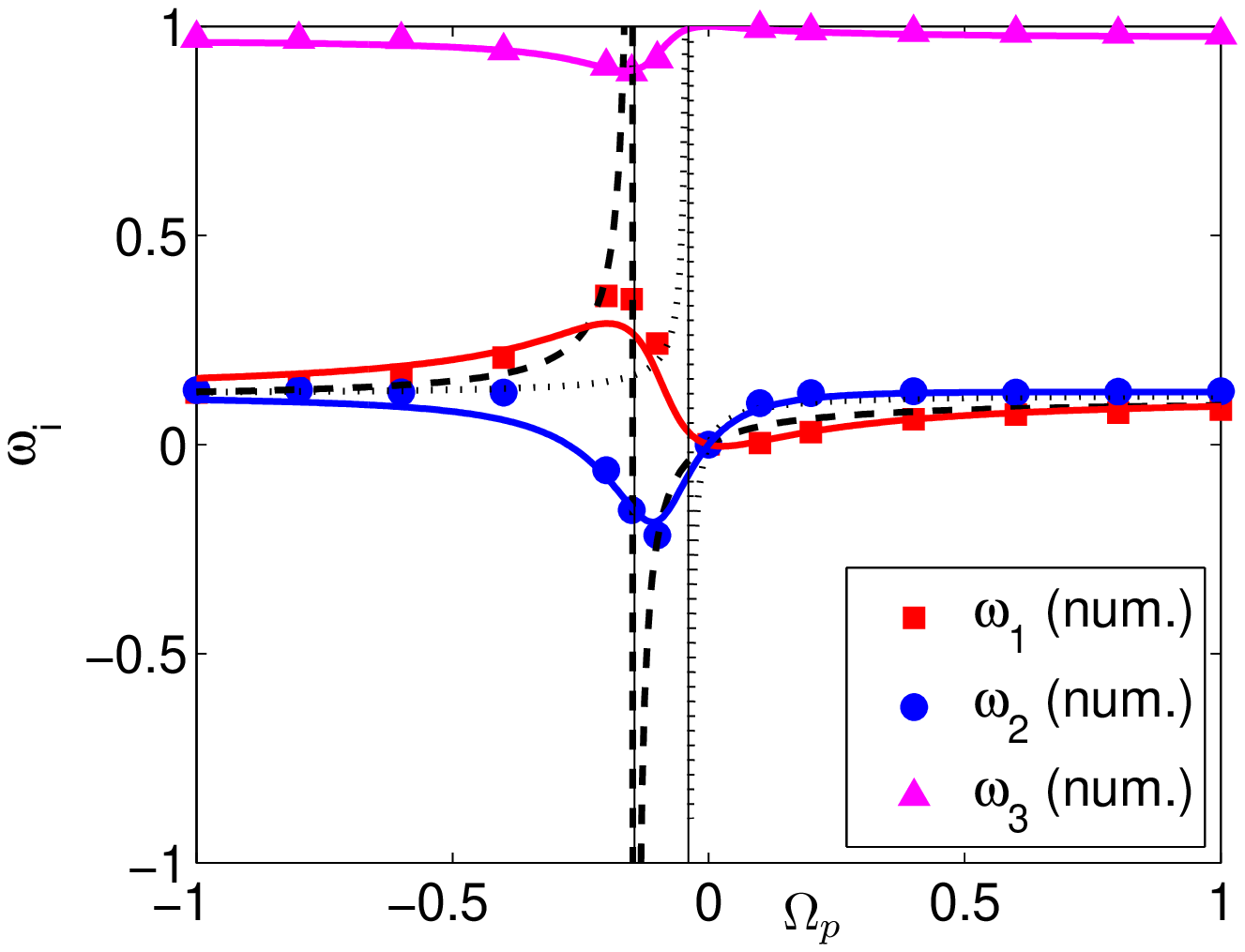}}
     \end{tabular}
    \caption{\it{
The rotation rate components of the flow in a spheroid, shown in
(a), are compared in figure (b) with those in the ellipsoid with the
same aspect ratio $a_3/a_1=0.86$ ($\theta=10^{\circ}$,
$\theta_2=45^{\circ}$ and $E=1/600$). The theoretical rotation rate
components of the inviscid Poincar\'e solution and of its extension
to triaxial ellipsoids are given by the black dashed and dotted
lines, assuming that $\omega_3=\beta_{12}=\frac{a_1^2+a_2^2}{2a_1
a_2}$ (see appendix \ref{instability}). The theoretical rotation
rate components of the Busse viscous solution and of its extension
to triaxial ellipsoids are represented by the continuous lines. The
two inviscid resonances $\gamma_1$ and $\gamma_2$ are represented by
the vertical black lines. (a) In the spheroid, the validity of our
numerical model is confirmed over a large range of precession rate
by its good agreement with Busse solution. (b) In the ellipsoid with
the same aspect ratio but with $\epsilon=0.1$, our analytical
solution is in good agreement with the numerical results.}}
    \label{cebronfig12}
  \end{center}
\end{figure}

A first series of simulations have been performed for
$a_3/a_1=0.86$, $\theta=10^{\circ}$, $\theta_2=45^{\circ}$,
$E=1/600$, $\epsilon=0.1$ and various precession rates. Results are
presented in figure \ref{cebronfig12} (b). An excellent agreement is
found between the numerical simulations and the analytical viscous
solution all along the explored range, which demonstrates the
validity of our extended viscous theory. Note also that the triaxial
Poincar\'e inviscid flow is recovered far from the resonances. As
already noted above, an important feature of the triaxial geometry
is the apparition of a second resonance. As already noted in the
section \ref{inviscid}, according to the equation
(\ref{eq:resonance}), the two resonances are in the retrograde
regime here because $a_1,a_2>a_3$. Note also that as already
described in the literature (see e.g. \cite{Noir_2003}), the
viscosity naturally smoothes the inviscid resonance peaks but also
modifies their position.

The tilt-over mode is thus well described by our analytical model at
relatively large Ekman number and relatively low precession rate and
tidal eccentricity. However, two types of instabilities can be
expected for more vigorous flows: the tidal or elliptical
instability at relatively small Ekman and/or large eccentricity
$\epsilon$, and the precession instability at relatively small Ekman
and/or large precession rate. Focusing on the tidal instability, we
keep the Ekman number above the precession instability threshold,
and we increase the eccentricity above the elliptical instability
threshold.

\begin{figure}                  % Chaque figure doit avoir pour nom nomfig1.eps,
  \begin{center}
     \begin{tabular}{ccc}
      \setlength{\epsfysize}{7.0cm}
      \subfigure[]{\epsfbox{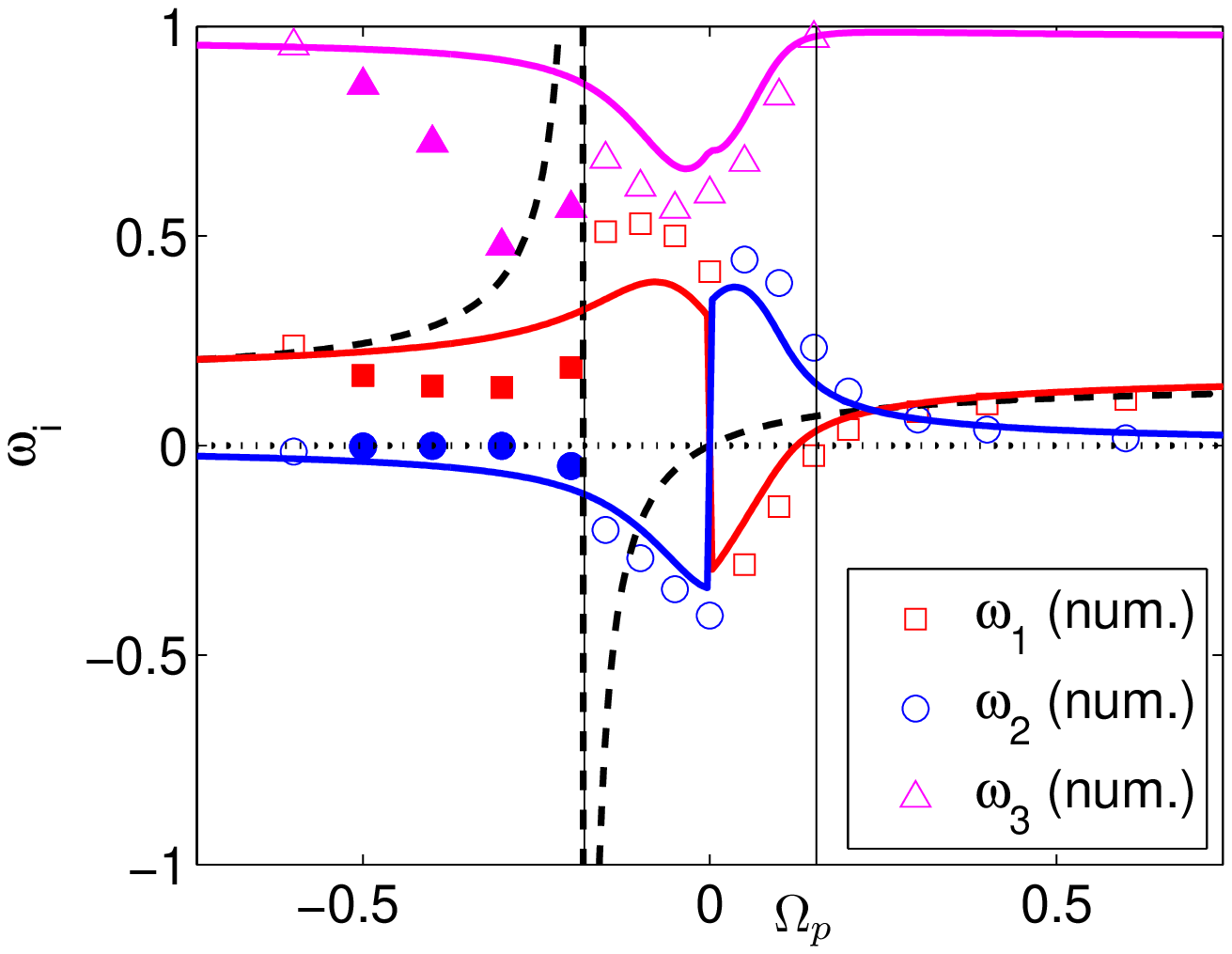}} \\
      \setlength{\epsfysize}{7.0cm}
      \subfigure[]{\epsfbox{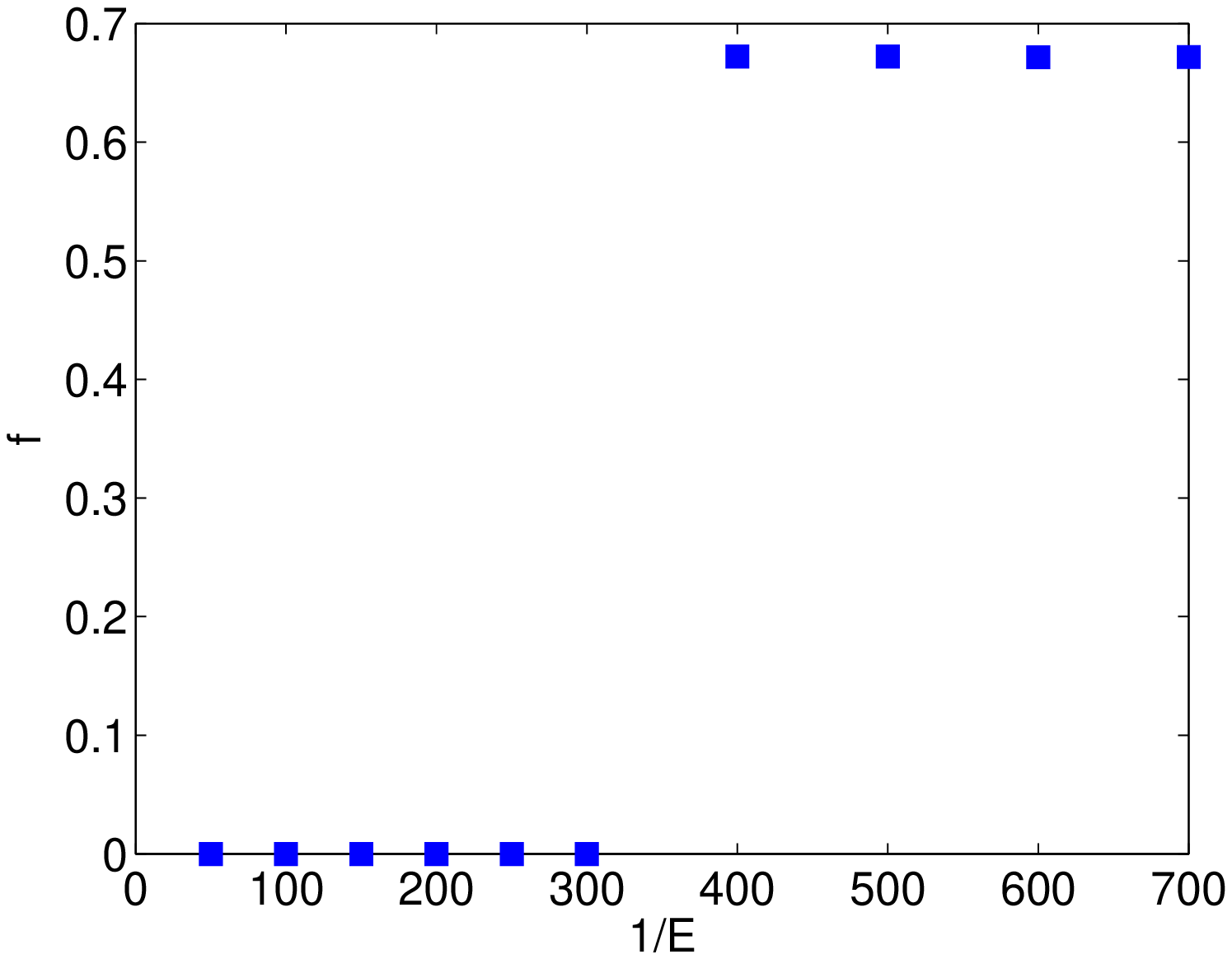}}
      \end{tabular}
   \caption{\it{
    We consider the case $\theta=10^{\circ}$, $a_3/a_1=0.86$, $\theta_2=0$ and $\epsilon=0.317$ (corresponding to $(a_1+a_2)/2=a_3$).
(a) Evolution of the rotation rate components of the flow with the
amplitude of precession for a fixed value of the Ekman number
$E=1/600$. Legend is the same as in figure \ref{cebronfig12} (b). Filled symbols correspond to cases where the flow is oscillatory whereas empty symbols corresponds to a stationary flow.
Note that the sign reversal at $\Omega_p=0$ is due to the two
possible symmetric orientations of the spinover mode of the tidal
instability. (b) We now focus on the $\Omega_p=-0.4$ case where the
flow is oscillatory at $E=1/600$ as shown in (a), and we study the
evolution of the pulsation $f$ of the flow with the Ekman number.
The flow is stationary for an Ekman number larger than $E=1/300$,
and then oscillates at a fixed pulsation $f\approx 0.67$.}}
    \label{cebronfig13}
  \end{center}
\end{figure}

The appearance and form of the tidal instability in the presence of
a rotating tidal deformation aligned with the rotation axis have
been studied elsewhere \cite[][]{LeBars09, Cebron_2010}. This case
corresponds in our notations to the specific configuration $\theta =
0$. The most well-known mode is the so called spin-over mode, which
appears for low values of tidal bulge rotation. As shown in appendix
\ref{instability}, a theoretical stability analysis demonstrates
that this mode may grow upon the Poincar\'e flow. Note that the
spin-over mode, which is an unstable mode of the tidal instability,
and the tilt-over mode, which is the base flow of precession, are
actually identical, hence should both be described by our stationary
analytical solution. This is indeed the case, as shown in figure
\ref{cebronfig13} (a) around $\Omega_p=0$ where numerical and
analytical solutions are in good agreement. Moreover, by comparison
with the known results of the elliptical instability in the special
case $\theta = 0$ (see \cite{LeBars09} and \cite{Cebron_2010} for
details), we expect the elliptical instability to lead to an
unstationary mode for $\Omega_p$ below the resonance band of the
spin-over, and to no elliptical instability for $\Omega_p$ above the
resonance band of the spin-over. These results are also recovered in
figure \ref{cebronfig13} (a), where oscillatory motions are indeed
observed for $\Omega_p \in [-0.5;-0.2]$, whereas good agreement is
found for $\Omega_p>0$ between analytical and numerical results.
Finally, a systematic study of the excited mode frequency as a
function of the Ekman number for $\Omega_p=-0.4$ is shown in figure
\ref{cebronfig13} (b). As expected from the stability analysis
presented in appendix \ref{instability}, the transition from a
stationary flow to an unstationary one is found below $E \approx
1/300$.

\section{Conclusion}

This paper presents the first analytical solution and the first
numerical simulations of the precessing flow inside a triaxial
ellipsoid. The extension of Busse viscous solution to this more
complex geometry is shown to provide an accurate description of the
stationary flow, which may be destabilized by oscillating modes of
the tidal and/or precession instabilities for more vigorous forcing.
Clearly, the complete study of instabilities in such a system
deserves more work and will be the subject of future studies with
increased numerical power and a new experimental set-up. But the
study of the stationary flow performed here already highlights the
fact that even a very small tidal deformation significantly modifies
the resulting flow. This conclusion is directly relevant to the
dynamics of planetary cores and atmospheres, where precession and
tidal deformation simultaneously take place.

\section{Appendix: Instability of the inviscid Poincar\'e flow in a triaxial ellipsoid}\label{instability}
The work \cite{Naing_2009} have recently studied the stability of a
rotating flow with elliptical streamlines in the particular case
where the rotation axis executes a constant precessional motion
about a perpendicular axis. Using the WKB method, \cite{Naing_2009}
achieve to quantify both the influence of this particular precession
on the elliptical instability, and the influence of the strain on
the Coriolis instability for such an unbounded cylindrical base
flow. In this appendix, we consider the stability of the inviscid
Poincar\'e flow in triaxial ellipsoids against small inviscid
perturbative rotations. The theoretical expression of the growth
rate for perturbations that are linear in space variables (i.e.
corresponding to the classical spin-over mode) can be readily
obtained with the same method as in \cite{Gledzer92}, but with the
base flow $\mathbf{U}$ obtained in section \ref{inviscid}, using an
arbitrary value of $\omega_3$. As remind in \cite{Kerswell_2002},
the most general perturbation-flow linear in the spatial coordinates
is
\begin{eqnarray}
u_k-U_k=  K_{i+1} (t)\ \frac{a_i}{a_{i-1}}\ x_{i-1} - K_{i-1}(t)\ \frac{a_i}{a_{i+1}}\ x_{i+1}
\end{eqnarray}
where the scalar amplitude $K_i$ of these small inviscid
perturbative rotations can be written $K_i=\varepsilon\ k_i\ e^{\sigma t}$. The total flow $\mathbf{u}$ have to satisfy the inviscid
vorticity equation. Then, $\mathbf{U}$ being solution flow at order
$0$, the solution flow at first order in $\epsilon$ is solution of
\begin{eqnarray}
M\ \mathbf{k}=0
\end{eqnarray}
where $\mathbf{k}=\left( k_i \right)_{i \in (1,2,3)}$ and $M$ is a $3 \times 3$ matrix given by

\begin{eqnarray}
M=
    \left [
   \begin{array}{ccc}
      A_{12} & -A_{21} & -B_{12}\ \sigma \\
      C_1 & B_{13}\ \sigma & D_{12} \\
      -B_{23}\ \sigma & C_2 & D_{21} \\
   \end{array}
   \right ]
\end{eqnarray}
where
 $$A_{ij}=2\ \Omega_{p,j}\ \frac{a_3}{a_j}
\frac{\gamma_i}{\gamma_j},$$
 $$B_{ij}=\frac{a_i^2+a_j^2}{a_i\ a_j},$$
$$C_{i}=\frac{\gamma_i}{a_3\ a_i},$$
 $$D_{ij}=-4 \frac{\Omega_{p,i}\
\Omega_{p,3}\ a_j^2}{\gamma_i}.$$

The solution flow at first order is non-trivial if $\det(M)=0$, and
the growth rate $\sigma$ is obtained in solving this equation.
Actually, the $3^{rd}$ degree equation $\det(M)=0 $ writes in the
simple form $\sigma^3 +p\ \sigma=0$, and the growth rate is given by:
\begin{eqnarray}
\sigma=\sqrt{-p}=\sqrt{\frac{A_{12} B_{13} D_{21}+A_{21} B_{23} D_{12}-C_1 C_2 B_{12}}{B_{13}B_{23}B_{12}}}
\end{eqnarray}

Note that the dimensionless base flow of \cite{Gledzer92} is
recovered for the particular value $\omega_3=\beta_{12}$. It can
also be noticed that the extension of the result of \cite{Gledzer92}
to the particular case of a rotating tidal bulge at
$\boldsymbol{\Omega_p}=(0,0,\Omega_{br})$ given in
\cite{Cebron_2010}, is recovered:
\begin{eqnarray}
\sigma=\sqrt{-\frac{C_1 C_2}{B_{13}B_{23}}}=\sqrt{-\frac{(a_1^2-a_3^2 + 2 a_1 a_2 \Omega_{br} )(a_2^2-a_3^2+2  a_1 a_2 \Omega_{br})}{(a_1^2+a_3^2)(a_2^2+a_3^2)}}.\label{eq:sigma_c2}
\end{eqnarray}
Note finally that this method gives the theoretical inviscid growth
rate, which has then to be corrected by a viscous surfacic damping
term $\chi \sqrt{E}$ \cite[][]{Kudlick}, where $\chi$ is a constant
$\chi \sim 2.62$. The threshold is then given explicitly by
\begin{eqnarray}
E_c=\frac{1}{\chi^2} \frac{A_{12} B_{13} D_{21}+A_{21} B_{23}
D_{12}-C_1 C_2 B_{12}}{B_{13}B_{23}B_{12}}
\end{eqnarray}

For the case presented in figure \ref{cebronfig13}, i.e.
$\theta=10^{\circ}$, $a_3=(a_1+a_2)/2$, $\theta_2=0$ and
$\epsilon=0.317$, we predict a critical Ekman number for instability
around $E_c \approx 1/300$, and for $E = 1/600$, we predict an
unstable spin-over mode in the range $\Omega_p \in [-0.13; 0.12]$.

% \bibliographystyle{ieeetr}
% \bibliography{biblio1}

\begin{thebibliography}{10}

\bibitem{Malkus93}
W.~Malkus, {\em Lectures on Solar and Planetary Dynamos (Energy sources for
  planetary dynamos)}.
\newblock Cambridge University Press, London, 1993.
\newblock (ed. MRE Proctor, AD Gilbert).

\bibitem{Tilgner_2005}
A.~Tilgner, ``Precession driven dynamo,'' {\em Phys. Fluids}, vol.~17,
  p.~034104, 2005.

\bibitem{Poincare}
R.~Poincar\'e, ``Sur la pr\'ecession des corps d\'eformables,'' {\em Bull.
  Astr. 27}, vol.~27, pp.~321--356, 1910.

\bibitem{Stewartson_1963}
K.~Stewartson and P.~Roberts, ``On the motion of a liquid in a spheroidal
  cavity of precessing rigid body,'' {\em J. Fluid Mech.}, vol.~17, pp.~1--20, 1963.

\bibitem{Busse}
F.~Busse, ``Steady fluid flow in a precessing spheroidal shell,'' {\em J. Fluid
  Mech.}, vol.~33, pp.~739--751, 1968.

\bibitem{Greenspan}
H.~Greenspan, {\em The theory of rotating fluids}.
\newblock Cambridge University Press, Cambridge, 1968.

\bibitem{malkus68}
W.~Malkus, ``Precession of the earth as the cause of geomagnetism,'' {\em
  Science}, vol.~160, pp.~259--264, 1968.

\bibitem{Vanyo_1995}
J.~P. Vanyo, P.~Wilde, P.~Cardin, and P.~Olson, ``Experiments on precessing
  flows in the earth's liquid core,'' {\em Geophys. J. Int.}, vol.~121,
  pp.~136--142, 1995.

\bibitem{Noir_2003}
J.~Noir, P.~Cardin, D.~Jault, and J.~P. Masson, ``Experimental evidence of
  nonlinear resonance effects between retrograde precession and the tilt-over
  mode within a spheroid,'' {\em Geophys. J. Int.}, vol.~154, pp.~407--416,
  2003.

\bibitem{Hollerbach_1995}
R.~Hollerbach and R.~R. Kerswell, ``Oscillatory internal shear layers in
  rotating an1d precessing flows,'' {\em J. Fluid Mech.}, vol.~298,
  pp.~327--339, 1995.

\bibitem{Quartapelle_1995}
L.~Quartapelle and M.~Verri, ``On the spectral solutions of the 3-dimensional
  navier-stokes equations in spherical and cylindrical regions,'' {\em Bull.
  Astr. 27}, vol.~90, pp.~1--43, 1995.

\bibitem{Tilgner_1999a}
A.~Tilgner, ``Magnetohydrodynamic flow in precessing spherical shells,'' {\em
  Geophys. J. Int.}, vol.~379, pp.~303--318, 1999.

\bibitem{Tilgner_2001}
A.~Tilgner and F.~H. Busse, ``Fluid flows in precessing spherical shells,''
  {\em J. Fluid Mech.}, vol.~426, pp.~387--396, 2001.

\bibitem{Noir_2001}
J.~Noir, D.~Jault, and C.~P., ``Numerical study of the motions within a slowly
  precessing sphere at low ekman number,'' {\em J. Fluid Mech.}, vol.~437,
  pp.~283--299, 2001.

\bibitem{Walker_1994}
M.~R. Walker and C.~F. Barenghi, ``High resolution numerical dynamos in the
  limit of a thin disk galaxy,'' {\em Geophys. Astrophys. Fluid Dyn.}, vol.~76,
  pp.~265--281, 1994.

\bibitem{Tilgner_1999b}
A.~Tilgner, ``Non-axisymmetric shear layers in precessing fluid ellipsoidal
  shells,'' {\em J. Fluid Mech}, vol.~136, pp.~629--636, 1999.

\bibitem{Lorenzani_2001}
S.~Lorenzani and A.~Tilgner, ``Fluid instabilities in precessing spheroidal
  cavities,'' {\em J. Fluid Mech.}, vol.~447, pp.~111--128, 2001.

\bibitem{Lorenzani_2003}
S.~Lorenzani and A.~Tilgner, ``Inertial instabilities of fluid flow in
  precessing spheroidal shells,'' {\em J. Fluid Mech.}, vol.~492, pp.~363--379,
  2003.

\bibitem{Schmitt_2004}
D.~Schmitt and D.~Jault, ``Numerical study of a rotating fluid in a spheroidal
  container,'' {\em J. Comp. Phys.}, vol.~197, pp.~671--685, 2004.

\bibitem{Wu_2009}
C.-C. Wu and P.~H. Roberts, ``On a dynamo driven by topographic precession,''
  {\em Geophys. Astrophys. Fluid Dyn.}, vol.~103, pp.~467--501, 2009.

\bibitem{KerswellMalkus}
R.~Kerswell and W.~V.~R. Malkus, ``Tidal instability as the source for {Io}'s
  magnetic signature,'' {\em Geophys. Res. Lett.}, vol.~25, pp.~603--606, 1998.

\bibitem{Lacaze_2004}
L.~Lacaze, P.~Le~Gal, and S.~Le~Diz\`{e}s, ``Elliptical instability in a
  rotating spheroid,'' {\em J. Fluid Mech.}, vol.~505, pp.~1--22, 2004.

\bibitem{LeBars09}
M.~Le~Bars, M.~Lacaze, S.~Le~Diz\`es, P.~Le~Gal, and M.~Rieutord, ``Tidal
  instability in stellar and planetary binary system,'' {\em Phys. Earth
  Planet. Int.}, vol.~178, pp.~48--55, 2010.

\bibitem{Cebron_2010}
D.~C\'ebron, M.~Le~Bars, J.~Leontini, P.~Maubert, and P.~Le~Gal, ``A systematic
  numerical study of the tidal instability in a rotating ellipsoid,'' {\em
  Phys. Earth Planet. Int.}, vol.~182, pp.~119--128, 2010.

\bibitem{Kerswell_2002}
R.~Kerswell, ``Elliptical instability,'' {\em Annu. Rev. Fluid Mech.}, vol.~34,
  pp.~83--113, 2002.

\bibitem{Waleffe_1990}
F.~A. Waleffe, ``On the three-dimensional instability of strained vortices,''
  {\em Phys. Fluids}, vol.~2, pp.~76--80, 1990.

\bibitem{LeBars07}
M.~Le~Bars, S.~Le~Diz\`es, and P.~Le~Gal, ``Coriolis effects on the elliptical
  instability in cylindrical and spherical rotating containers,'' {\em J. Fluid
  Mech.}, vol.~585, pp.~323--342, 2007.

\bibitem{Kerswell_1993}
R.~Kerswell, ``The instability of precessing flow,'' {\em Geophys. Astrophys.
  Fluid Dyn}, vol.~72, pp.~107--144, 1993.

\bibitem{Kerswell_1995}
R.~Kerswell, ``On the internal shear layers spawned by the critical regions in
  oscillatory ekman boundary layers,'' {\em J. Fluid Mech.}, vol.~298,
  pp.~311--325, 1995.

\bibitem{Lagrange2008}
R.~Lagrange, P.~Meunier, C.~Eloy, and F.~Nadal, ``Instability of a fluid inside
  a precessing cylinder,'' {\em Phys. Fluids}, vol.~20, p.~081701, 2008.

\bibitem{Noir_2000}
J.~Noir, {\em Ecoulement d'un fluide dans une cavit\'e en pr\'ecession:
  approches num\'eriques et exp\'erimentales}.
\newblock PhD thesis, Universit\'e Joseph-Fourier, Grenoble 1, 2000.

\bibitem{Owen_1989}
J.~M. Owen and R.~H. Rogers, {\em Flow and heat transfer in rotating disc
  systems (Vol. 1, Rotor-stator systems)}.
\newblock Research Studies Press, Taunton, UK \& John Wiley, NY, 1989.

\bibitem{Meunier2008}
P.~Meunier, C.~Eloy, R.~Lagrange, and F.~Nadal, ``A rotating fluid cylinder
  subject to weak precession,'' {\em J. Fluid Mech.}, vol.~599, pp.~405--440,
  2008.

\bibitem{Naing_2009}
M.~M. Naing and Y.~Fukumoto, ``Local instability of an elliptical flow
  subjected to a coriolis force,'' {\em J. Phys. Soc. Jpn.}, vol.~78,
  p.~124401, 2009.

\bibitem{Gledzer92}
E.~B. Gledzer and V.~M. Ponomarev, ``Instability of bounded flows with
  elliptical streamlines,'' {\em J. Fluid Mech}, vol.~240, pp.~1--30, 1992.

\bibitem{Kudlick}
M.~Kudlick, {\em On the transient motions in a contained rotating fluid}.
\newblock PhD thesis, Massachusetts Institute of Technology, 1966.

\end{thebibliography}

\end{document}